\documentclass[9pt,twocolumn,twoside]{opticajnl}
\journal{opticajournal} 

\setboolean{shortarticle}{true}

\usepackage{amsmath,amssymb}

\usepackage{graphicx}
\usepackage{dcolumn}
\usepackage{bm}
\usepackage{mathtools}
\usepackage{color}

\DeclareMathOperator{\Tr}{Tr}

\DeclarePairedDelimiterX{\abs}[1]{\vert}{\vert}{#1}
\DeclarePairedDelimiterX{\norm}[1]{\lVert}{\rVert}{#1}
\DeclarePairedDelimiterX{\expval}[1]{\langle}{\rangle}{#1}
\DeclarePairedDelimiterX{\ket}[1]{\vert}{\rangle}{#1}
\DeclarePairedDelimiterX{\bra}[1]{\langle}{\vert}{#1}
\DeclarePairedDelimiterX{\innerproduct}[2]{\langle}{\rangle}{#1\delimsize\vert\mathopen{}#2}
\DeclarePairedDelimiterX{\outerproduct}[2]{\vert}{\vert}{#1\delimsize\rangle\!\delimsize\langle\mathopen{}#2}
\DeclarePairedDelimiterX{\mel}[3]{\langle}{\rangle}%
{#1\delimsize\vert\mathopen{}#2\delimsize\vert\mathopen{}#3}

\usepackage{lineno}

\title{Detecting changes to sub-diffraction objects with quantum-optimal speed and accuracy}

\author[1,*]{Michael R Grace}
\author[2]{Saikat Guha}
\author[1]{Zachary Dutton}

\affil[1]{Physical Sciences and Systems, Raytheon BBN, Cambridge, MA 02138, USA}
\affil[2]{James C. Wyant College of Optical Sciences, University of Arizona, Tucson, AZ 85721, USA}

\affil[*]{michael.grace@rtx.com}

\begin{abstract} 
Detecting if and when objects change is difficult in passive sub-diffraction imaging of dynamic scenes. We consider the best possible tradeoff between responsivity and accuracy for detecting a change from one arbitrary object model to another in the context of sub-diffraction incoherent imaging. We analytically evaluate the best possible average latency, for a fixed false alarm rate, optimizing over all physically allowed measurements of the optical field collected by a finite 2D aperture. We find that direct focal-plane detection of the incident optical intensity achieves sub-optimal detection latencies compared to the best possible average latency, but that a three-mode spatial-mode demultiplexing measurement---in concert with on-line statistical processing using the well-known CUSUM algorithm---achieves this quantum limit for sub-diffraction objects. We verify these results via Monte Carlo simulation of the change detection procedure and quantify a growing gap between the conventional and quantum-optimal receivers as the objects are more and more diffraction-limited.
\end{abstract}

\begin{document}

\maketitle

	Object dynamics hold a wealth of desirable information in live-sample biological microscopy, remote sensing, space situational awareness, and surveillance. In particular, discrete changes in a target object from one spatial configuration to another often necessitate immediate actions such as changes in sample illumination or strategic decisions. For many remote monitoring or object tracking needs, passive imaging is the only feasible modality by which to detect and respond to critical changes in the object(s) being monitored. 
	
	Unfortunately, as with many imaging tasks, the physical principle of diffraction can impede responsiveness to dynamics that manifest at small length scales relative to the system point spread function (PSF).	When objects are obfuscated by the PSF due to diffraction from a finite aperture, detecting even the simplest dynamical processes, such as an instantaneous change from one object configuration to another, becomes challenging with conventional imaging methods. As an example, Ref.~\cite{boettiger_super-resolution_2016} required \emph{in situ} labeling with photoswitchable fluorescent biomarkers, live-cell superresolution microscopy protocols, and post-processing detection analyses to reveal transitions between conformational states of chromatin folding. The challenges posed by diffraction-dominated imaging conditions are especially severe in ``on-line" change detection settings when the task is to not only identify \emph{whether} a change occurs but also \emph{when} it occurs in real time. In this case, the goal of the imaging procedure is to trigger on a change in the object as soon as possible after the changepoint, i.e., with the lowest latency, while minimizing inaccuracy in the form of ``false-alarm" (FA) detections, i.e., triggering before a true change occurred~\cite{lorden_procedures_1971}.
	
	A potential avenue to improve the speed and accuracy of on-line change detection has recently arisen in the form of ``quantum-inspired" superresolution imaging methods. By modeling the optical field collected from a scene as a quantum state, one can compute ultimate bounds on the information that can be extracted from the light~\cite{tsang_quantum_2016}. These quantum analyses have identified a number of fully classical pre-detection spatial transformations that measure the light in a spatial basis with a quantum-optimal signal to noise ratio, enhancing imaging performance over that exhibited by a direct focal plane imager. These spatial mode demultiplexing (SPADE) receivers have been applied in simulations and experiments to demonstrate quantitative enhancements in sub-diffraction parameter estimation~\cite{tsang_quantum_2016,tsang_semiparametric_2019}, hypothesis testing~\cite{lu_quantum-optimal_2018,grace_identifying_2022,zhang_super-resolution_2020}, and scene reconstruction~\cite{bearne_confocal_2021,frank_passive_2023}. Could such approaches yield similar advantages for detecting sub-diffraction changes in scene monitoring scenarios with existing technology?

\begin{figure}[t]
		\centering
		\includegraphics[width=.45\textwidth]{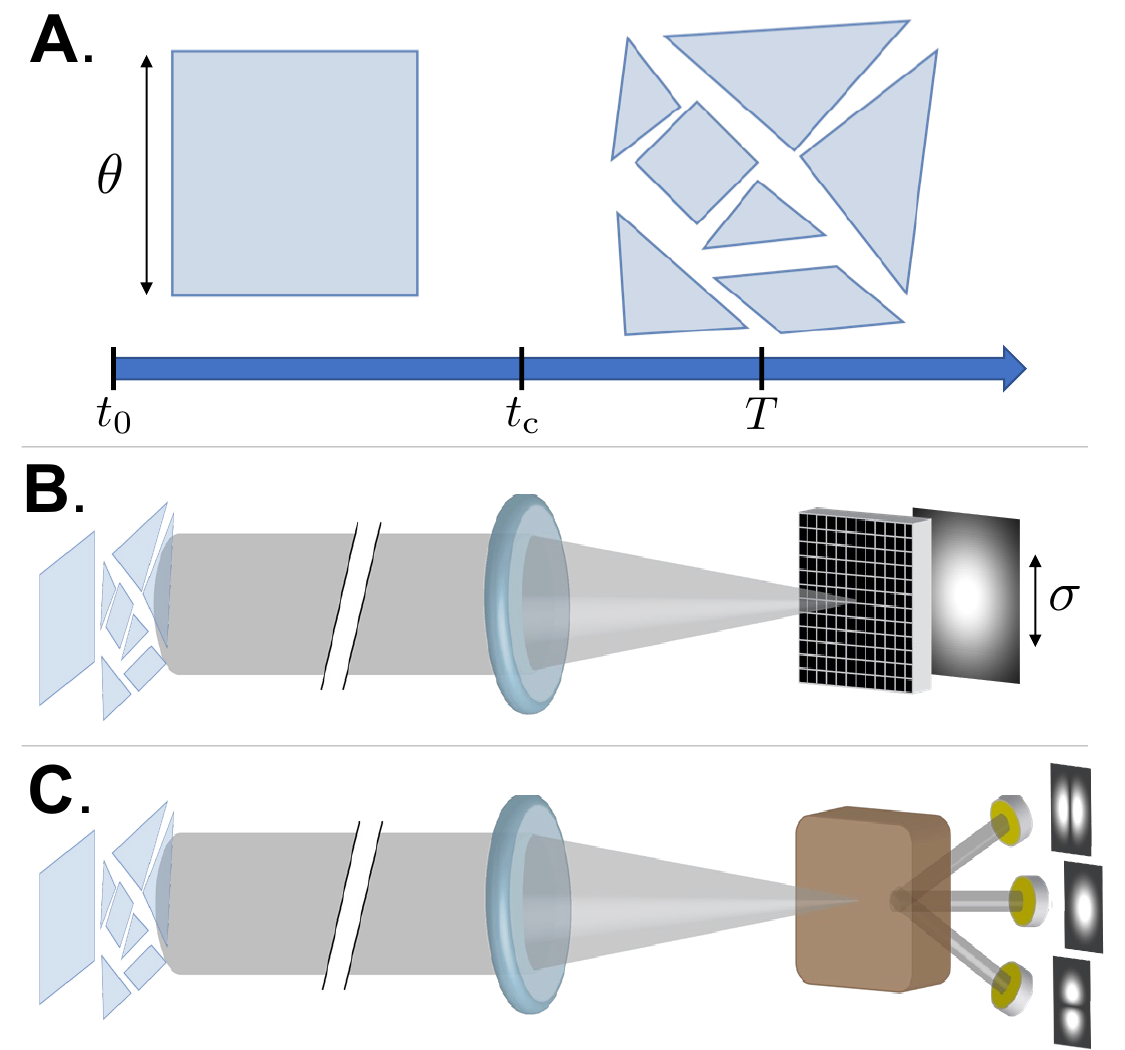}
		\caption{\textbf{A}. Example objects before and after a change at time $t_c$. \textbf{B}. Direct imaging generates a diffraction-blurred image of the object. \textbf{C}. TriSPADE sorts three orthogonal spatial modes of the collected light to individual detectors. }
		\label{fig:Task}
	\end{figure}

	In our Letter, we show the potential of quantum-inspired imaging for sub-diffraction change detection via an analytical derivation of the ultimate quantum limit on generalized detection latency as well as Monte Carlo simulations of real-time imaging. We first consider the best possible tradeoff between speed and accuracy, among all passive imaging schemes allowed by the laws of quantum mechanics, for detecting a change at time $t_{\rm c}$ from one arbitrary 2D object to another (Fig.~\ref{fig:Task}A.). We assume the detection procedure is tracked in discrete time steps that each consists of one or more measurements performed on a tensor product quantum state $\eta_j^{\otimes M}$ that models the received optical field over $M$ orthogonal temporal modes accumulated during the $t^{\rm th}$ time step, where $j=1$ if $t\leq t_c$ and $j=2$ if $t>t_c$. We seek to evaluate any given combination of optical measurement and post-processing scheme for both its responsiveness to real-time changes and its accuracy in avoiding erroneous detections. Our metric of choice for detection speed is the average latency $\bar{\tau}\equiv\mathbb{E}_{t_{\rm c}}(\tau)$, where the expectation value is taken over the possible measurement records produced by a change at time $t_c$ and where latency of a given imaging trial that triggers at time $T$ is given by $\tau\equiv T-t_{\rm c}$. We evaluate detection accuracy using the mean time to false alarm $\bar{T}_{\rm FA}\equiv\mathbb{E}_{\infty}(T)$, which quantifies how many pre-change time steps the receiver will run on average before incorrectly triggering.

	The optimal balance between low latency (low $\bar{\tau}$) and high accuracy (high $\bar{T}_{\rm FA}$) was recently derived in Ref.~\cite{fanizza_qusum_2022} for on-line change detection between a generalized pair of quantum states. Optimizing over all physical measurements and post-processing schemes results in~\cite{fanizza_qusum_2022}
	\vspace{-3pt}
	\begin{equation}
		\bar{\tau}^\ast=\frac{\ln(\bar{T}_{\rm FA})}{M S(\eta_2\vert\vert\eta_1)},
		\label{eq:QuantumLimit}
	\end{equation}
	where $\bar{\tau}\geq\bar{\tau}^\ast$ is an ultimate bound on the worst-case mean latency (i.e., conditioned on the most unfavorable pre-change measurement record) and where $S(\eta_2\vert\vert\eta_1)\equiv\Tr[\eta_2\ln(\eta_2/\eta_1)]$ is the quantum relative entropy (QRE) between the two states in a single temporal mode. This ultimate performance tradeoff is achievable with some quantum-optimal measurement scheme yielding sequential measurement outcomes $z_t$ with likelihood functions $p(z_t\vert\eta_j^{\otimes M})$ for hypotheses $j\in[1,2]$. 
	
		From \eqref{eq:QuantumLimit}, it is clear that analyzing the quantum limited tradeoff for change detection speed and accuracy amounts to computing the QRE between the two quantum states associated with the pre- and post-change object models in the imaging scenario. Following the methods of Ref.~\cite{grace_identifying_2022}, we calculate the QRE in the Poisson noise limit for \emph{any} two incoherent, 2D object models $m_j(\vec{x})$ that are severely blurred by an arbitrary 2D coherent PSF $\psi(\vec{x})$. Since natural thermal sources often exhibit $\epsilon\ll1$ mean photons per coherence time, such that the possibility of coherent multiphoton detection can be ignored, we use the weak-source approximation $\eta_j=(1-\epsilon)\outerproduct{0}{0}+\epsilon\rho_j+O(\epsilon^2)$, where $\outerproduct{0}{0}$ is a vacuum state and $\rho_j$ is a reduced density operator conditioned on detecting one photon~\cite{tsang_quantum_2016}. Since vacuum is uninformative with respect to state discrimination, the denominator in \eqref{eq:QuantumLimit} becomes $N S(\rho_2\vert\vert\rho_1)$, where $N=\epsilon M$ is the mean photon number per time step. In the basis of single-photon eigenkets $\ket{\vec{x}}$ corresponding to image-plane photon arrival positions $\vec{x}$, the quasimonochromatic single-photon quantum states are given by~\cite{tsang_subdiffraction_2017}
	\begin{equation}
		\rho_j=\iint_{-\infty}^{\infty}\frac{1}{\mu^2}m_j\left(\frac{\vec{x}}{\mu}\right)\outerproduct{\psi_{\vec{x}}}{\psi_{\vec{x}}}d^2\vec{x},
		\label{eq:rhoj}
	\end{equation}
	where $\ket{\psi_{\vec{x}}}=\iint_{-\infty}^{\infty}\psi(\vec{a}-\vec{x})\ket{\vec{a}}d^2\vec{a}$ encodes the effect of the aperture and where $\mu$ is the imaging system magnification.
	
	As in Ref.~\cite{grace_identifying_2022}, we define the parameter $\gamma=\mu\theta/\sigma$ that quantifies the magnification-scaled ratio between the largest spatial extent $\theta$ among the candidate objects and the characteristic PSF width $\sigma$. This allows us to define  non-dimensionalized versions of the object models $\tilde{m}_j(\vec{x})=\theta^2m_j(\theta\vec{x})$, PSF $\tilde{\psi}(\vec{x})=\sigma\psi(\sigma\vec{x})$, and PSF autocorrelation function $\tilde{\Gamma}(\vec{x})=\Gamma(\sigma\vec{x})$, where $\Gamma(\vec{x})=\iint_{-\infty}^{\infty}\psi^{\ast}(\vec{a})\psi(\vec{a}-\vec{x})d^2{\vec{a}}$. These nondimensionalized quantities capture spatial properties of the objects or PSF independent of the object-PSF ratio $\gamma$, allowing us to independently focus our analytic results in the $\gamma\ll1$ regime corresponding to the sub-diffraction limit. In this regime, we showed in Ref.~\cite{grace_identifying_2022} that considering terms only up to $O(\gamma^2)$ allowed us to truncate the density matrices $\rho_j$ to a small number of dimensions in the PSF adapted (PAD) basis, which is constructed via a spatial mode orthogonalization starting from the 2D PSF itself~\cite{kerviche_fundamental_2017,rehacek_optimal_2017}. Assuming that the 2D spatial centroid of both objects is registered to the origin at the imaging plane, the state can be written as a small perturbation away from the pure state $\ket{\psi_{\vec{\Omega}}}$ corresponding to a point source at the origin of the object plane~\cite{grace_identifying_2022}. Using our result for the QRE between perturbed quantum states~\cite{grace_perturbation_2021}, we find 
	\vspace{-4pt}
	\begin{equation}
     \begin{aligned}
		S(\rho_2\vert\vert\rho_1)=&\Bigg(\left[m_{1,x^2}-m_{2,x^2}+m_{2,x^2}\ln\left(\frac{m_{2,x^2}}{m_{1,x^2}}\right)\right]\Gamma_{x^2}\\
        &+\left[m_{1,y^2}-m_{2,y^2}+m_{2,y^2}\ln\left(\frac{m_{2,y^2}}{m_{1,y^2}}\right)\right]\Gamma_{y^2}\Bigg)\gamma^2\\
        &+O(\gamma^3),
		\label{eq:QRE}
    \end{aligned}
	\end{equation}
	where $m_{j,x^ky^l}=\iint_{-\infty}^{\infty}x^ky^l\tilde{m}_j(\vec{x})d^2\vec{x}$ and $\Gamma_{x^ky^l}= -[\Re(\partial^{k+l}\tilde{\Gamma}(\vec{x})/\partial x^k\partial y^l)]_{\vec{x}=\{0,0\}}$ are spatial moments of the object models and derivatives of the 2D autocorrelation $\tilde{\Gamma}(\vec{x})=\Gamma(\sigma\vec{x})$ of the PSF. This expression is the foundational result of our paper and shows that there must be a measurement such that the relative entropy between the outcomes observed from any two sub-diffraction objects, and therefore also the detection latency for detecting changes between them, depends quadratically on the object-PSF ratio $\gamma$. We note that this result can also be directly applied to asymmetric hypothesis testing, implying that the optimal exponential decay rate of the type II error when testing between any two sub-diffraction objects must be $O(\gamma^2)$ and is given by \eqref{eq:QRE} according to the quantum Stein's lemma~\cite{hiai_proper_1991,ogawa_strong_2000}.
	
\begin{figure}[t]
		\centering
		\includegraphics[width=.42\textwidth]{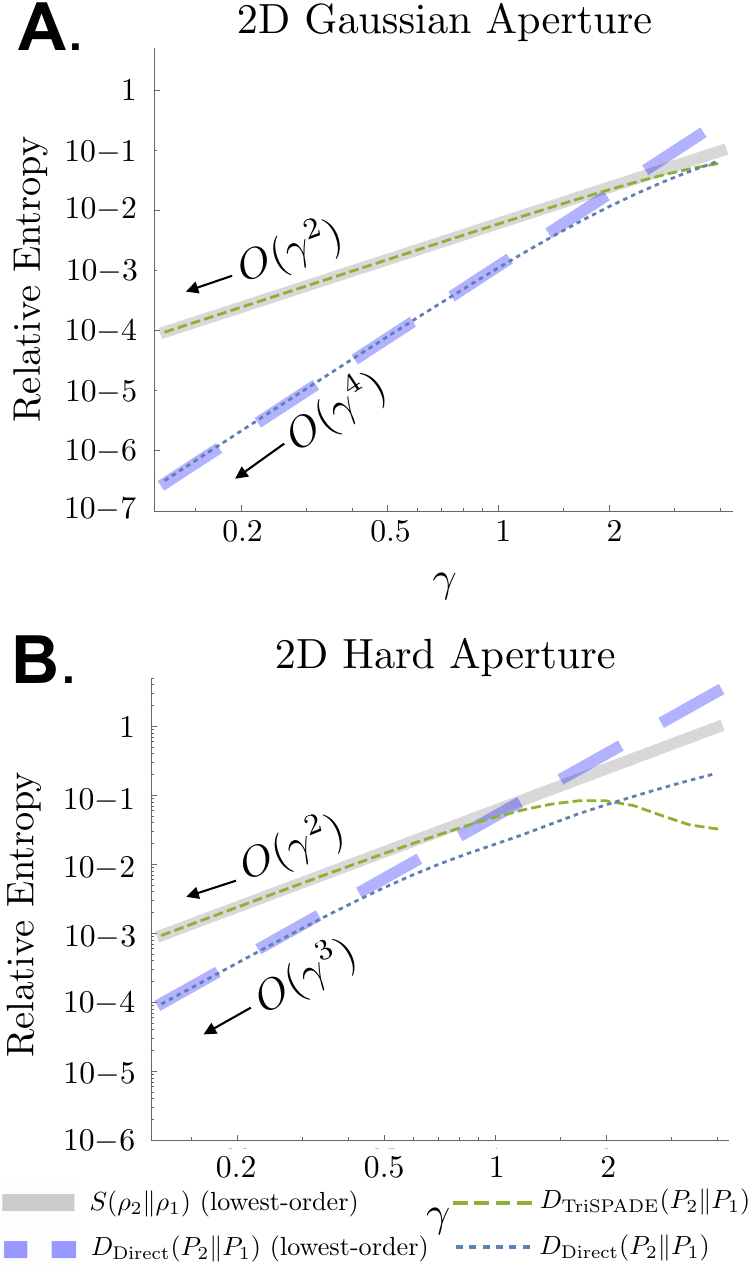}
		\caption{Quantum relative entropy compared with classical relative entropies of direct imaging and TriSPADE for the two objects represented in Fig.~\ref{fig:Task}A. The gap between the RE of direct imaging and the QRE in the sub-diffraction limit ($\gamma\ll1$) is of order $O(\gamma^2)$ for a 2D Gaussian aperture with a zeroless PSF but is only of order $O(\gamma)$ for the 2D hard aperture with an Airy disk PSF. }
		\label{fig:RelativeEntropy}
	\end{figure}

	We would like to compare the performance of particular measurement schemes to the analytical quantum limit in order to find an optimal measurement. The latency/false alarm tradeoff for a particular measurement is known to obey the inequality~\cite{lorden_procedures_1971}
	\begin{equation}
		\bar{\tau}\geq \frac{\ln(\bar{T}_{\rm FA})}{M D(P_2\vert\vert P_1)},
		\label{eq:ClassicalBound} 
	\end{equation}
	where $P_j\equiv p(z_t\vert\eta_j)$ give the outcome probabilities of each i.i.d. measurement of a copy of the quantum state $\eta_j$ and where $D(P_2\vert\vert P_1)\equiv \int P_2 \log(P_2/P_1) dz$ is the classical relative entropy (RE) between outcome probabilities of the two object models. For any imaging scenario with a PSF that is zero nowhere in $\mathbb{R}^2$, we find that the RE for a direct imaging measurement that uses an idealized focal plane array of intensity detectors (Fig.~\ref{fig:Task}B.)  is given by 
	\begin{equation}
    \begin{aligned}
		D_{\rm Direct}(P_2\Vert P_1)=&\frac{1}{8}[(m_{2,x^2}-m_{1,x^2})^2\Lambda_{x^2}\\
        &+(m_{2,y^2}-m_{1,y^2})^2\Lambda_{y^2}]\gamma^4+O(\gamma^5)
		\label{eq:RE}
    \end{aligned}
	\end{equation}
	which exhibits the fourth-order scaling $O(\gamma^4)$. This means that the detection latency achieved by a direct imaging measurement will necessarily be larger than the quantum limit by a factor that increases as the scene becomes more diffraction-limited. On the other hand, we find that the RE of a ``TriSPADE" measurement that sorts the first-order PAD-basis spatial modes in two orthogonal directions against the zeroth order PSF-matched mode (Fig.~\ref{fig:Task}C.)~\cite{grace_identifying_2022} equals the QRE up to second order in $\gamma$. TriSPADE is therefore a quantum-optimal measurement for sub-diffraction change detection, exhibiting a substantially greater relative entropy than direct imaging in the sub-diffraction regime. In Fig.~\ref{fig:RelativeEntropy}A. we plot the numerically computed REs for direct imaging and TriSPADE against the lowest-order analytical results for the QRE [\eqref{eq:QRE}] and direct imaging RE [\eqref{eq:RE}] for a 2D Gaussian aperture with coherent PSF $\psi(\vec{x})=1/\sqrt{2\pi\sigma^2}\exp[-\vec{x}\cdot\vec{x}/(4\sigma^2)]$ to visualize the quadratic gap between the quantum limit and direct imaging. Interestingly, for PSFs with zeros, such as the Airy disk PSF $\psi(\vec{x})=j_1(\pi\sqrt{\vec{x}\cdot\vec{x}}/\sigma)/(\sqrt{\pi\vec{x}\cdot\vec{x}})$ arising from a hard circular aperture, we find numerically that the RE for direct imaging exhibits an improved scaling of $O(\gamma^3)$ when $\gamma\ll1$. This observation is reminiscent of that from Ref.~\cite{paur_reading_2019} for estimating the separation between two sub-Rayleigh point sources with direct imaging and a hard aperture, and it results in a linear scaling gap with respect to $\gamma$ compared to the quantum limit (Fig.~\ref{fig:RelativeEntropy}B.).

    To verify our results, we simulated the change detection procedure for a representative on-line change detection scenario: a square that shatters into disordered pieces (Fig.~\ref{fig:Task}A.). The optimal post-processing approach for on-line change detection is known to be the CUSUM algorithm~\cite{lorden_procedures_1971}, which computes a running cumulative sum of the log-likelihood ratios $G_t=[G_{t-1}+\ln(P_2/P_1)]^{+}$, where $[u]^{+}=0$ if $u\leq0$. The algorithm stops and declares a change has occurred at the first time $T$ for which $G_T>h$, where $h$ is a user-defined threshold (see Fig.~\ref{fig:CUSUMResults}A.). The mean latency for the algorithm is thus $\bar{\tau}=\mathbb{E}_{t_{\rm c}}[T]-t_{\rm c}$.
	
	An example imaging trial with $\gamma=0.25$ and a mean photon number per time step of $N=500$ (Fig.~\ref{fig:CUSUMResults}A.) illustrates the operation of the CUSUM algorithm for sub-diffraction imaging. We observe that TriSPADE is much more sensitive to the object change event than direct imaging, triggering in this example with a latency of $\tau_{\rm TriSPADE}=3$ vs. $\tau_{\rm Direct}=26$. Our chosen CUSUM threshold of $h=\ln(25000)=10.13$ ensures a probability of false alarm $P_{\rm FA}\leq0.001$ over 25 pre-change time steps~\cite{fanizza_qusum_2022}.  We first checked that we obtain the expected behavior for different CUSUM thresholds $h$ when using TriSPADE. Across the domain $0<\gamma<1$ we confirmed that the average latency obeys the relation {$\bar{\tau}=(h+\mathbb{E}_0[x])/[N S(\rho_2\vert\vert\rho_1)]$ for $h\gg0$ (Fig.~\ref{fig:CUSUMResults}B.) and that the average time to a false alarm is lower bounded by $\bar{T}_{\rm FA}\geq e^h/\mathbb{E}_{\infty}[e^{-x}]$ (Fig.~\ref{fig:CUSUMResults}C.), where $x$ is the amount by which the cumulative log-likelihood ratio overshoots the threshold~\cite{lorden_procedures_1971,fanizza_qusum_2022}.
	
	We report the ensemble results of our Monte Carlo simulation in Fig.~\ref{fig:CUSUMResults}D., where we find that the information-favorable data from the TriSPADE measurement enables mean latencies that saturate the quantum limit $\bar{\tau}^{\ast}$, as computed using Eqs.~\eqref{eq:QuantumLimit} and \eqref{eq:QRE}, in the deeply sub-diffraction regime $\gamma\ll1$. On the other hand, we fit the results obtained by direct imaging to a trend line whose latency/false-alarm tradeoff exhibits sub-optimal scaling $O(\gamma^3)$ for sub-diffraction imaging. When the imaged object is much smaller than the PSF, e.g., $\gamma=0.125$, we find that TriSPADE reduces the mean latency by a factor of 10 compared to direct imaging, while for larger objects the latencies are comparable. 
	
	Our results demonstrate that quantum-inspired, classical measurements can enhance responsivity to changes in sub-diffraction objects, promising near-term benefits for high spatio-temporal resolution of dynamic scenes. One nuance in our results is that the pre- and post-change object models must be co-located such that their 2D spatial centroids are exactly identical for our results to hold, and the advantage over direct imaging disappears if the object change involves a relative shift in object location. We argue that this seemingly strong restriction is frequently irrelevant in practice, since in many scenarios the object will be dynamically diffusing or traveling and space and will need to have its location tracked in real time alongside monitoring for change detection, so a change-induced shift in centroid will be indistinguishable and the two candidate objects will both be modeled with the identical pre-estimated centroid. Our proposed TriSPADE receiver can be implemented using mature pre-detection optical mode sorters, which have been shown to enable passive sub-diffraction imaging capabilities and are fully modular and compatible with most modern imaging systems, simply replacing the camera sensor at the back focal plane. The three-detector TriSPADE design has the additional benefit of measurement compression, suppressing the effect of excess detector noise and greatly reducing the computational overhead that is often the limiting factor for real-time image processing.

    \begin{figure}[t]
		\centering
		\includegraphics[width=.49\textwidth]{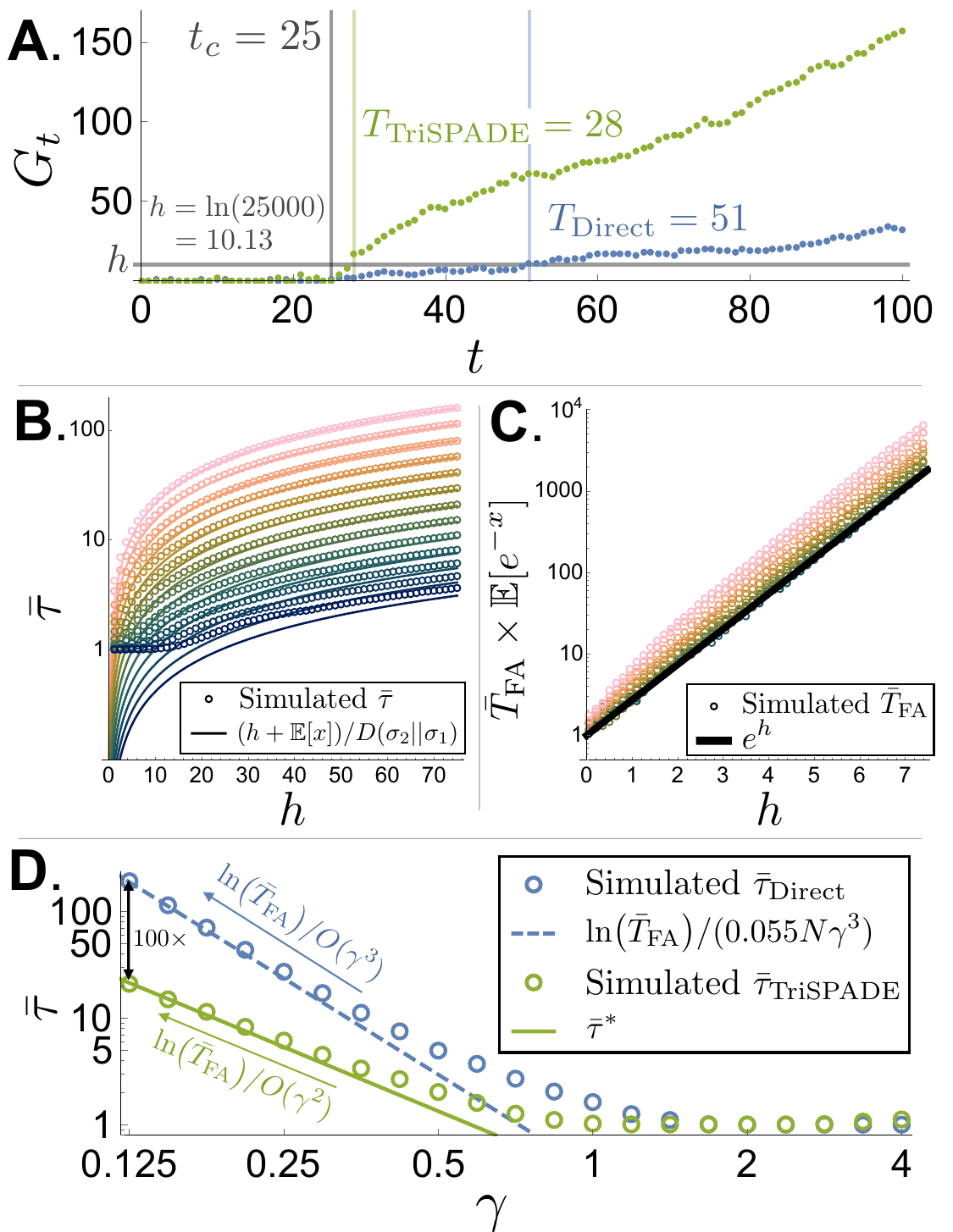}
		\caption{\textbf{A}. Example CUSUM run for direct imaging and TriSPADE. We set $N=500$ for all simulations. \textbf{B}. Mean latency and \textbf{C}. mean time to false alarm for TriSPADE (light to dark: $\gamma\in[0.125,1]$ in ascending factors of $\sqrt[4]{2})$, as functions of the CUSUM threshold $h$, compared with theoretical predictions. \textbf{D}. Mean CUSUM latency for direct imaging and TriSPADE vs. the quantum limit $\bar{\tau}^{\ast}$. }
		\label{fig:CUSUMResults}
	\end{figure}
 
\begin{backmatter}
\bmsection{Funding}
SG acknowledges valuable discussions with Ravi Tandon about the CUSUM algorithm. This research was supported by the DARPA IAMBIC Program under Contract No. HR00112090128. The views, opinions and/or findings expressed are those of the authors and should not be interpreted as representing the official views or policies of the Department of Defense or the U.S. Government.

\bmsection{Supplemental document}
See Supplement 1 for supporting content. 

\end{backmatter}


\bibliography{CUSUM}

\end{document}